\documentclass[10pt,letterpaper]{article}
\usepackage[utf8]{inputenc}
\usepackage[T1]{fontenc}
\usepackage{amsmath}
\usepackage{amsfonts}
\usepackage{amssymb}
\usepackage{graphicx}
\author{James S. Graber}
\title{JYx2}
\begin{document}
	\begin{center}
		{\huge A New Rotating Metric \\
		that Generalizes \\ \medskip
		the Yilmaz Exponential Metric} \\
		\bigskip \bigskip
\end{center}\bigskip \begin{center}
	{\huge James S. Graber \\
	jgraber@mailaps.org \\ \smallskip
	jimgraber@gmail.com}
	\end{center}
	
	\pagebreak
	In this paper, we present a new rotating metric that generalizes the non-rotating Yilmaz Exponential metric [Yilmaz (1958, 1972)], in the same way that the Kerr metric generalizes the Schwarzschild metric [Kerr (1963), Schwarzschild (1916)].
	\smallskip
	
	This new rotating metric is derived from the Yilmaz metric by a recent variation of the Newman-Janis (1972) technique, due to Dadhich (2013).
	\medskip
	
	Dadhich’s technique depends heavily on the hypothesis of ellipsoidal symmetry, which is most easily represented in the following so-called standard coordinate system, here expressed without loss of generality as:
	\bigskip
	\begin{equation}
	ds^2	=	A(R)\: dt^2 -B(R)\: dR^2 -R^2\: d\theta^2  - R^2\: sin^2 \theta\: d\phi^2	.
	\end{equation}
	
		\bigskip
	
	The Yilmaz exponential metric is usually expressed in isotropic coordinates as
		\bigskip
	\begin{equation}
		ds^2 = e^{[-2 M/r]} dt^2 - e^{[2 M/r]} ( dr^2 + r^2 d\theta^2 + r^2 sin^2 \theta d\phi ^2)
		\end{equation}	
		
	\bigskip
	                                                                   Therefore, our first step is to convert the Yilmaz metric into the standard form. Inspecting the coordinates of $d\theta$ and $d\phi$ makes it clear that

	\begin{equation}
R^2 = r^2 e^{[2 M/r]} .	
	\end{equation}

	Hence, 
		
	\begin{equation}
	R = r e^{[M/r]}		
	\end{equation}

	Using Mathematica to solve $R = r e^{[M/r]}$ for $r$, we obtain

	\begin{equation}
		r = -M / ProductLog[ -M/R]	
	\end{equation}

	Substituting this result back into the isotropic Yilmaz equation above, we obtain the following representation of the Yilmaz metric in standard coordinates:

	\begin{equation}
		\begin{split}
			ds^2 = M^2 / ( R^2 ProductLog[-M/R]^2 ) dt^2 - dR^2 /(1+ ProductLog[-M/R])^2 \\
		-R^2 d\theta^2 -R^2 sin^2 \theta d\phi ^2
	\end{split}
	\end{equation}

	Next we take advantage of the recent work of Dadhich(2013), where he develops the following two master equations to convert from a spherical non-rotating metric to an ellipsoidal rotating metric, based on the use of standard form coordinates and on the hypothesis of ellipsoidal symmetry. In these equations, $a$ is the rotation parameter (dimensionless angular momentum) and is also the measure of ellipticity. $M\,a$ is the angular momentum.
	
	\pagebreak
	Dadhich Equation 3 (Dadhich 2013):

	\begin{equation}
	ds^2 =  A(dt – a sin^2\theta d\phi)^2  - dr^2 /A-\rho^2 d\theta^2 - sin^2\theta [(r^2+a^2) d\phi- a dt]^2/ \rho^2		
	\end{equation}

	where  $\rho^2 = r^2 + a^2 cos^2 \theta$ and $A=(r^2+a^2)/ \rho^2$
	
	\bigskip 	
	
	Dadhich Equation 5 (Dadhich 2013):

	\begin{equation}
		\begin{split}
	ds^2 = (FFF(r)/\rho^2)(dt - a sin^2\theta d\phi)^2 - dr^2 (\rho^2/GGG(r)) - \rho^2 d\theta^2 \\- sin^2\theta [(r^2+a^2) d\phi- a dt]^2/ \rho^2
\end{split}		
	\end{equation}

	Combining these two equations, we obtain:

	\begin{equation}
		\begin{split}
	ds^2 = (FFF(r)(r^2+a^2)/\rho^2)(dt - a sin^2\theta d\phi)^2 - dr^2 (\rho^2/((r^2 + a^2)GGG(r)) \\- \rho^2 d\theta^2 - sin^2\theta [(r^2+a^2) d\phi- a dt]^2/ \rho ^2
	\end{split}		
	\end{equation}

	Using the matching coordinates for $dt^2$ and $dr^2$ from the standard form Yilmaz metric above, we substitute

	\begin{equation}
		FFF(r) = M^2/(r^2 ProductLog[-M/r]^2)	
	\end{equation}

 and

	\begin{equation}
	GGG(r) = (1+ ProductLog[-M/r])^2	
	\end{equation}

	The result is this new rotating metric, which reduces to the Yilmaz metric when the rotation parameter $a$ is set to zero. 
		\bigskip
\begin{align*}		
	ds^2 = &M^2/(r^2 ProductLog[-M/r]^2)(r^2+a^2)
	(dt –a sin^2\theta d\phi)^2/ (r^2 + a^2 cos^2 \theta) \\
	&- sin^2\theta ((r^2+a^2) d\phi- a dt)^2/ (r^2 + a^2 cos^2 \theta) \\&
	- dr^2(r^2 + a^2 cos^2 \theta) 
	/((r^2 + a^2)(1+ ProductLog[-M/r])^2) \\ &
	- (r^2 + a^2 cos^2 \theta) d\theta^2
\end{align*}
	
	\pagebreak
	\bigskip
	[1] Dadhich, Naresh (2013) A novel derivation of the rotating black hole metric Gen Relativ. Grav. 45: 2383-2388
		\smallskip
		
	[2] Newman, E. T. and Janis, A. (1965) Note on the Kerr spinning-particle metric
	I. J. Math. Phys. 6, 915
		\smallskip
		
	[3] Yilmaz H (1958) New approach to General Relativity Phys. Rev. 111 1417-26
		\smallskip
		
	[4] Yilmaz H (1972) New theory of gravitation Nuovo Cimento B, Vol 10B, Ser. 11, p 79-101
		\smallskip
		
	[5] Schwarzschild, K. (1916) Über das Gravitationsfeld eines Massenpunktes nach der Einsteinschen Theorie
	Sitz. Konig. Preuss. Akad. Wiss. 7: 189-196
		\smallskip
		
	[6] Kerr, Roy P. (1963) Gravitational Field of a Spinning Mass as an Example of Algebraically Special Metrics
	Phys. Rev. Let. 11(5):237-238

		\bigskip

\end{document}